\documentstyle[11pt,newpasp,twoside,epsf]{article}

\newcommand\etal{{\rm et al}.\ }
\def\h2{\rm H_2}
\newbox\grsign \setbox\grsign=\hbox{$>$} \newdimen\grdimen \grdimen=\ht\grsign
\newbox\simlessbox \newbox\simgreatbox
\setbox\simgreatbox=\hbox{\raise.5ex\hbox{$>$}\llap
     {\lower.5ex\hbox{$\sim$}}}\ht1=\grdimen\dp1=0pt
\setbox\simlessbox=\hbox{\raise.5ex\hbox{$<$}\llap
     {\lower.5ex\hbox{$\sim$}}}\ht2=\grdimen\dp2=0pt

\newcommand\simless{\mathrel{\copy\simlessbox}}

\begin{document}

\title{Testing the Paradigm of Low-Mass Star Formation}
\author{Lee Hartmann}
\affil{Smithsonian Astrophysical Observatory, 60 Garden St., Cambridge,
MA 02138}

\begin{abstract}
Protostellar core formation is probably much more dynamic, and
magnetic fields are probably much less important, than has
been previously assumed in the standard model of low-mass star formation.
This revised picture has important consequences: it is
easier to understand the observed rapidity of star formation in
molecular clouds; cores are more likely to have structures favoring
high infall rates at early times, helping to explain the differences
between Class 0 and Class I protostars; and core structure and asymmetry
will strongly favor post-collapse fragmentation into binary and
multiple stellar systems. 
\end{abstract}

\section{Introduction}
What do we mean when we speak of ``testing the paradigm of low-mass star formation''?  
If it is simply questioning the sequence of events -- a molecular cloud core 
collapses at near free-fall velocities to a protostar plus disk system, 
followed by accretion through a disk onto the star,
and eventual planet formation (e.g., Shu, Adams, \& Lizano 1987)
-- then there is little controversy.
On the other hand, if one is testing details
from older models -- for example, 
that molecular cloud cores must lose substantial magnetic
flux by ambipolar diffusion before collapse; that cores are essentially
static before collapsing; and that core collapse occurs ``inside-out'--
then there is much less agreement.  Indeed, developments over the last
several years have cast significant doubt on many of these details.

More broadly, it is obvious that any theory of low-mass star formation
must encompass the end points: specifically, the formation 
of protostellar cores, and the formation of binary and/or
multiple systems (probably through post-collapse fragmentation 
in many if not most cases).  The standard paradigm is essentially
silent on these issues; it must be expanded if we are to understand
aspects of star formation such as the initial mass function.
  
In this contribution I will focus on the origin of protostellar cores, 
which has implications for core structure and therefore for subsequent 
phases of collapse and fragmentation.

\section{Molecular clouds and magnetic fields}

Because cores are subunits of molecular clouds, some attention must be paid
to the processes of cloud formation and dispersal.
For some time it was thought that giant molecular clouds 
are relatively long-lived (e.g., Solomon \etal 1979). 
This view implies that molecular clouds must be supported by 
supersonic turbulence (Norman \& Silk 1980; Larson 1981); that this turbulence is
responsible for preventing wide-spread gravitational collapse,
which would result in far too large a galactic star formation rate
(Zuckerman \& Evans 1974); and that cloud turbulence must be strongly
Alfvenic in character, to avoid dissipating too rapidly
(Arons \& Max 1975; see discussion in Shu, Adams, \& Lizano 1987).

A preliminary step toward changing this long-life view was taken
by Blitz \& Shu (1980).  They used several arguments to argue that
typical cloud lifetimes are $\simless 30$~Myr, including the concentration
of OB associations to spiral arms (implying that molecular gas
does not drift for a long time past the arms) and the rapidity with which massive
stars can disperse molecular gas. 

\begin{table}
\begin{center}
\begin{tabular}{c c c}
\tableline
\multicolumn{3}{c}{Table 1} \\
\multicolumn{3}{c}{Star forming regions} \\
\tableline
{Region} & {$<t>$~(Myr)} & {Molecular gas?} \\
\tableline
Coalsack   & --    & yes    \\
Cha III    & ?     & yes    \\
Orion Nebula&  1   & yes    \\
Taurus      &  2   & yes    \\
Oph         &  1   & yes    \\
Cha I,II    &  2   & yes    \\
Lupus       &  2   & yes    \\
MBM 12A     &  2   & yes    \\
IC 348      &  1-3 & yes    \\
NGC 2264    &  3   & yes    \\
Upper Sco   &  2-5   & no   \\
Sco OB2     &  5-15 & no    \\
TWA         &  $\sim$~10   & no    \\
$\eta$~ Cha     & $\sim$~10   & no    \\
\tableline
\end{tabular}
\end{center}
\end{table}

However, stellar population
ages suggest that cloud lifetimes in the solar neighborhood are
of order 3-5~Myr, i.e.\ an order of magnitude smaller than the Blitz-Shu estimate.
As shown in Table 1, taken from Hartmann, Ballesteros-Paredes, \& Bergin
(2001; HBB), it is very rare to find substantial molecular clouds without
at least some young stars forming within them.  In almost all
local clouds, the typical age of the stellar population is $\sim 1-3$~Myr.
And older groups, whether OB associations or low-mass groups, of 
ages 5-10 Myr have no associated molecular gas.  These data 
clearly demonstrate that star formation proceeds almost 
immediately upon cloud formation; that molecular clouds 
are transient structures; and that clouds are rapidly dispersed.

This picture has a number of important implications.
Numerical simulations have indicated that, contrary to expectation,
MHD turbulence decays rapidly (Stone, Ostriker, \& Gammie 1998; 
Mac Low \etal 1998; Mac Low 1999).  However, even with this rapid
dissipation, short cloud lifetimes comparable to or less than
crossing times (Elmegreen 2000; HBB) imply that  
turbulence need not be regenerated, but could simply be
left over from cloud formation.  This avoids difficulties using stellar
energy sources to maintain turbulence for long periods of
time, which in at least in the case of massive stars are much
more likely to disrupt the cloud than stabilize it.
Short cloud lifetimes also imply that the low rate of star 
formation is the result of reduced efficiency of conversion to 
gas to stars (Hartmann 1998; Elmegreen 2000) rather than 
slow cloud contraction as ambipolar diffusion proceeds.  

The short lag time between cloud formation and the onset of star formation
makes it unlikely that ambipolar diffusion can operate for a long
time.  In turn, this implies that large amounts of magnetic flux
need not be removed from the cloud; i.e., that the cores cannot
be highly magnetically subcritical.  This seems to fly in the
face of the so-called magnetic flux problem (Mestel \& Spitzer 1956),
which would seem to imply that formation of clouds from diffuse
interstellar conditions would result in far too much magnetic flux
for clouds of thousands of solar masses or less to contract without 
flux loss.  

To avoid the magnetic flux problem, supercritical clouds can be formed
by flows {\em along} the magnetic field (e.g., Mestel 1985). 
This requires large-scale flows, at least if molecular clouds
are to be formed by accumulation out of the diffuse galactic interstellar
medium. The accumulation length $l$ needed to produce a critical
cloud, measured along the magnetic field, adopting interstellar
medium parameters typical of the solar neighborhood, is
\begin{equation}
l_c ~\sim~  430 \, (B/5 \mu G) \, (n_H/1 {\rm cm}^{-3})\, {\rm pc}\,,
\end{equation}
While $l_c$ appears to be rather large, Blitz \& Shu (1980) considered
formation of giant molecular clouds from accumulation over 500 pc along
the field.  

Observations of large GMCs suggest that cloud accumulation lengths often are
very large.  In a number of cases, the lateral crossing
times of star-forming regions or young associations are substantially longer
than the ages of the stellar population (HBB).  It is difficult if not 
impossible to explain this coordination over large distance scales by propagating
information along the cloud.  The natural explanation is that the clouds
are swept up by large scale flows driven by either stellar winds, supernovae,
spiral density wave shocks, or a combination of all three.  The flows 
must have substantial correlation lengths to explain some of the largest
regions.  For instance, the Sco OB2 association has a projected extent of
order 150 pc (and an internal velocity dispersion in the plane of the sky
of only $1.5 {\rm km \,s^{-1}}$ or less; de Bruijne 1999).  Flows of hundreds
of pc in length are much more likely to exhibit such lateral coherence.

\begin{figure}[ht]
\plotfiddle{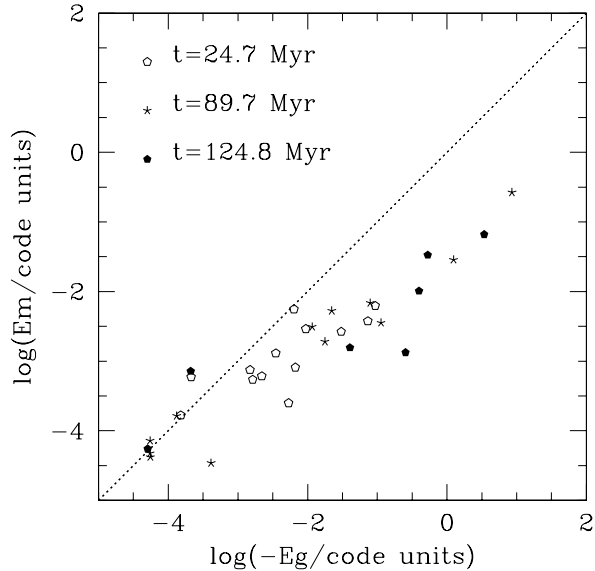}{8.cm}{0}{50}{50}{-150}{-70}
\caption{Cloud magnetic energies (vertical axis) vs. gravitational
energies (horizontal axis) for clouds in the simulations discussed in HBB.
The dotted line indicates magnetic criticality; the massive clouds are
supercritical.  From HBB.}
\end{figure}

The numerical simulations of the interstellar medium by 
Enrique Vazquez and his collaborators (Passot, Vazquez-Semadeni 
\& Pouquet 1995; see also Ballesteros-Paredes \etal 1999
and HBB) also suggest that accumulation lengths are large, and lead
to supercritical clouds.  
These ideal MHD simulations have some limitations -- they are two-dimensional,
and do not reach high densities -- but they are extremely important because they
cover very large regions (1 kpc on a side) and can thus address cloud formation
by flows on larger scales than any other simulation.

These simulations show that flows do extend for hundreds of pc, and
that they do tend to accumulate mass along magnetic field lines, as required
to make supercritical clouds out of the diffuse medium.  The simulations
also indicate that the flows can bend the magnetic field lines, if necessary,
to accomodate mass accumulation.  Finally, the simulations indicate that 
massive regions tend to be magnetically supercritical, as shown in Figure 1.

There is a fairly straightforward argument to be made why molecular
clouds formed in this way should be supercritical, as discussed in HBB.
In essence, dense regions tend to be magnetically-supercritical because
supercritical conditions are dynamically favorable for cloud formation.
If the internal magnetic field in a cloud creates a pressure 
force stronger than the force of gravity (i.e., the cloud is subcritical),
then the cloud will expand unless external pressures confine it.
Therefore dense clouds tend to have magnetic pressure forces weaker
than the external (ram) pressures.  But star formation generally proceeds in
clouds where gravitational forces are important in comparison with 
external pressure forces.  Thus, self-gravitating clouds tend to
have gravitational forces larger than both external pressure
and internal magnetic forces, i.e. they are supercritical.  
In addition, molecular clouds
only form in the solar neighborhood when the column density of material
is sufficient (generally, visual extinctions of order unity or larger)
to shield the molecules from the dissociating interstellar radiation field.
At this point, the self-gravity of the cloud is comparable to or larger
than that of typical interstellar medium pressures (Franco \& Cox 1986).
Thus when clouds accumulate enough material to become molecular, they
also tend to be both self-gravitating and supercritical.

Independent arguments suggesting that giant molecular clouds are at least critical
if not supercritical were given by McKee (1989).  Nakano (1998) also
made somewhat similar arguments that cores are generally supercritical.
Finally, recent observational studies of dense regions indeed suggest 
that cores are roughly critical if not supercritical (Crutcher 1999; Bourke \etal 2001).  

Note that, given the low efficiency of star formation on a global
scale (i.e., the entire molecular cloud), gravitational collapse need
only occur rapidly over limited scales.  Thus, even if the cloud as whole
is close to critical, some regions will be magnetically-strong and others
magnetically-weak, and the B-weak regions will probably collapse first.
Ambipolar diffusion is thus unlikely to slow protostellar collapse,
and magnetic forces do not prevent local rapid star formation
upon the formation of molecular gas, as required by observations.

\section{Filaments and fragmentation}

The rapidity of star formation in molecular clouds indicates the need for
a dynamic theory of protostellar core formation.  Indeed,
it is difficult to see how it could be otherwise, as one must take
diffuse molecular gas and concentrate it somehow, with some kind
of motions.  There are two extreme possibilities here.  One is that
the turbulent motions present in the cloud occasionally focus material
into small enough volumes that become Jeans unstable and then collapse
(e.g., Padoan \& Nordlund 1999).  The other extreme picture invokes
gravitational focusing in a relatively quiescent medium (e.g., Larson 1985).
It seems likely that the true solution lies somewhere in between these
two extremes; clouds are supersonically turbulent, but gravity must
obviously play an important role as well.  

As an initial step towards addressing these problems, consider the situation in
the Taurus star-forming region.  Taurus may not be typical of such
regions; it is of relatively low density, it is not making massive
stars, and it appears to have a somewhat distinctive initial
mass function in comparison with other regions (e.g., Brice\~no
\etal 2002).  However, it is close, easy to study,
and relatively quiescent.  If the paradigm of low-mass star
formation applies anywhere, it should apply in Taurus.

\begin{figure}[ht]
\plotfiddle{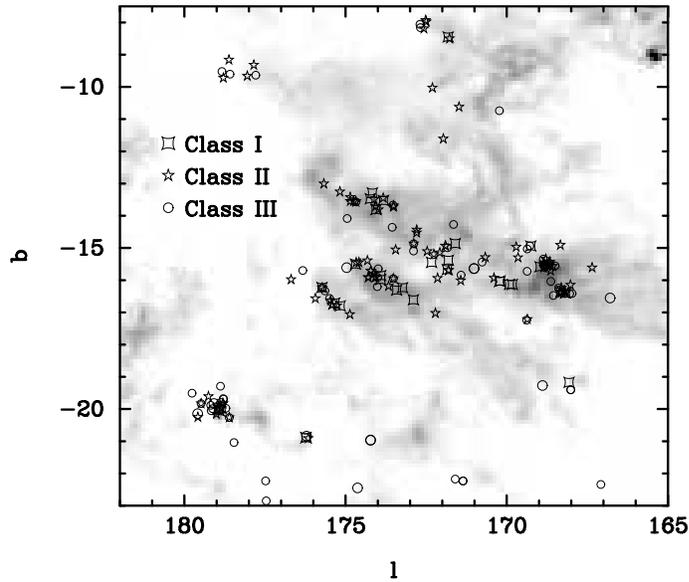}{8.cm}{0}{60}{60}{-160}{-100}
%\plotone{hartmann_l_fig2.eps}
\caption{Young stellar objects in the Taurus region, labelled by
their spectral energy distribution class, superimposed
upon the $^{12}$CO map of Megeath, Dame, \& Thaddeus (2002,
personal communication)}
\end{figure}

While it has long been recognized that the distribution of dense
gas in Taurus is filamentary, it has only recently been emphasized
that the distribution of young stars is extremely filamentary (Hartmann
2002).  In particular, as shown in Figure 2, on a large scale most of the
stars (and the dense gas) are distributed in three extensive, roughly
parallel bands.  One of these bands extends the entire length of the 
association.  

To see the implications of this structure, it is again useful to compare
with the results of numerical simulations.  Klessen (2001) and Klessen
\& Burkert (2001) have
computed dynamical structures with forced driving of turbulence on a variety
of scales.  Driving with fluctuations on small scales unsurprisingly
produces small-scale structures with no hint of large-scale filaments.
In contrast, low-spatial-wavenumber (large-scale) driving of turbulence
yielded long filaments in the Klessen et al. simulations.  Therefore
the Taurus structure suggests large-scale driving.  It is tempting to
identify this large-scale driving source with the large-scale flow
that I suspect formed Taurus in the first place.

\begin{figure}[ht]
\plottwo{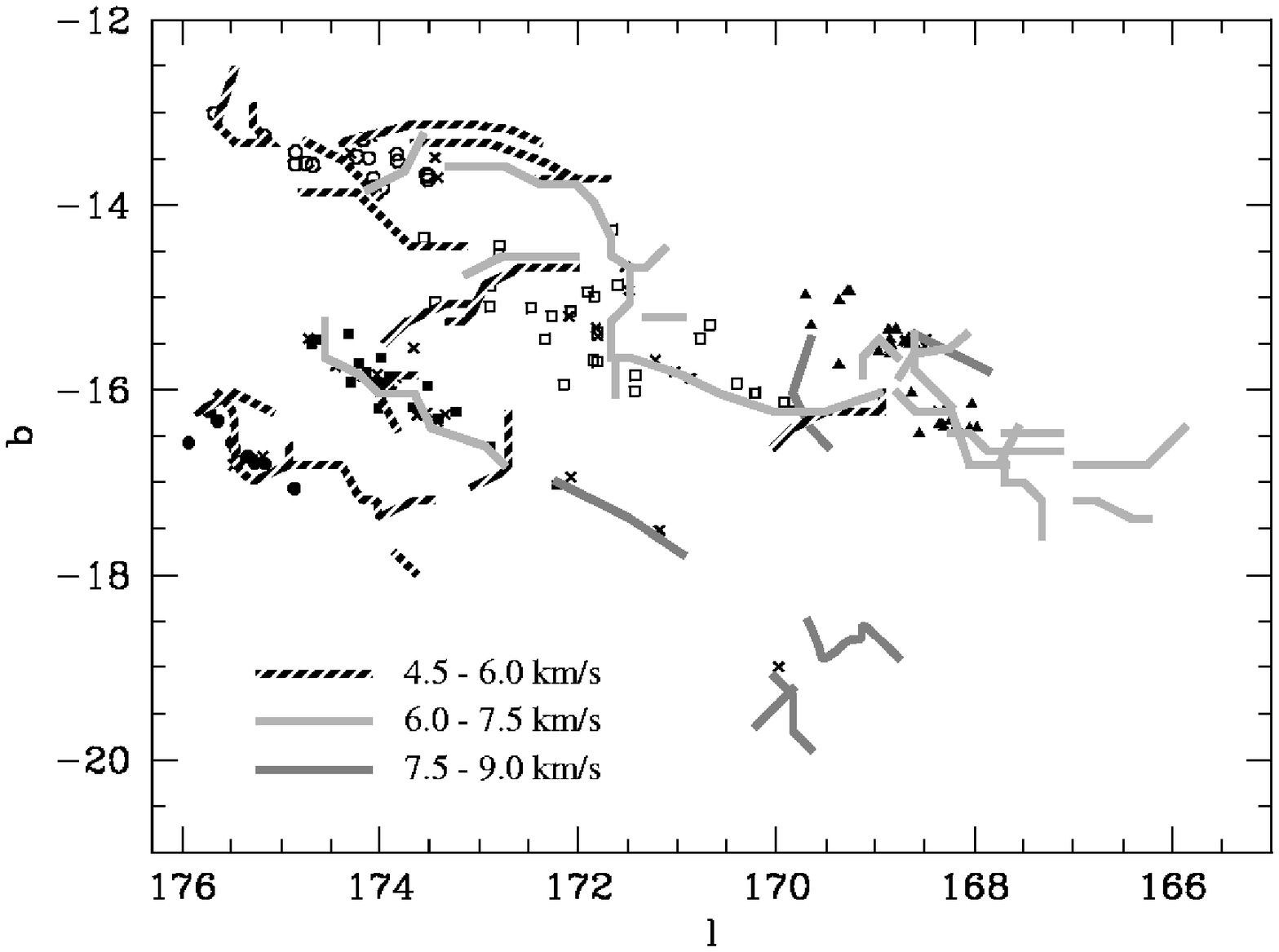}{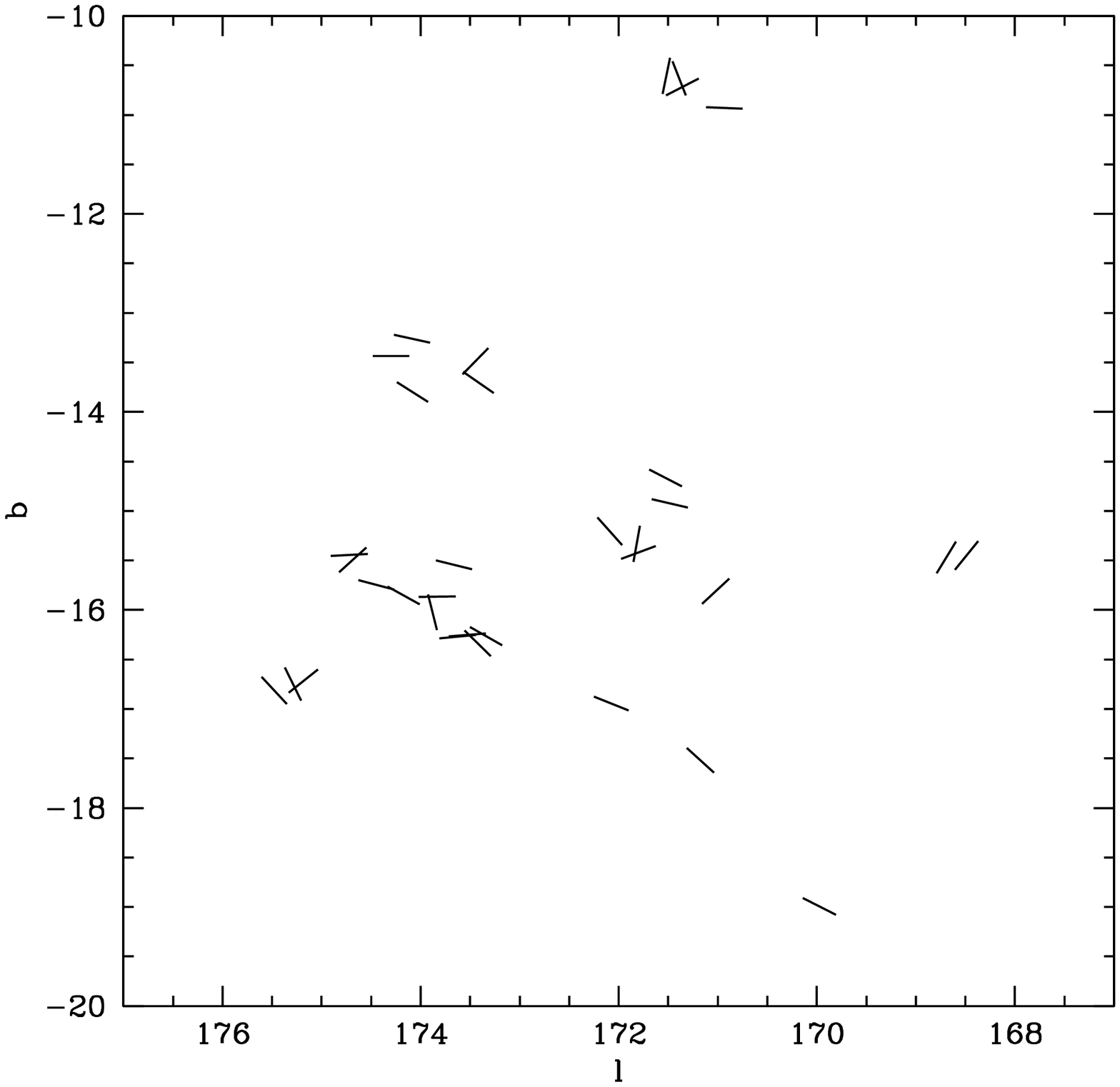}
\caption{Left: Approximate position of $^{13}$CO filaments in
the central region of Taurus found by Mizuno \etal (1995),
with young stars indicated.  The CO filaments
are sorted into the same LSR velocity ranges as Mizuno \etal.
Right: Spatial distribution of optical cores from Lee \& Myers (1999),
with the orientation of the line indicating the position angle of
the core major axis. Comparison of the two panels shows that
cores in Taurus are generally elongated along filaments.
From Hartmann (2002).}
\end{figure}

How are the filaments produced?  One possibiity is that turbulence is
entirely responsible, as in the Klessen et al. simulations.  There is
clearly smaller-scale structure in the dense gas filaments, as shown
in the left panel of Figure 3, which almost certainly reflects small-scale
motions of some kind.  Alternatively, gravity could play an important
role in focusing material into filaments, as for example in the simulations
by Miyama \etal (1987a,b).  

However the filaments are formed, it seems most likely to me that core
production is the result of gravitational fragmentation in the filaments,
as envisaged by Larson (1985).  It seems unlikely to me that even if
non-gravitational flows form the filaments, sub-flows would then arise
to break up the filament; in contrast, gravitational fragmentation should
naturally operate.

The gravitational fragmentation hypothesis predicts that the initial
scale of fragmentation should be longer than the typical width of the
filament.  In support of this hypothesis, as shown in the right hand
panel of Figure 3, cores are typically elongated along their host
filaments.
Note that this picture of gravitational fragmentation can only work
if the filaments are supercritical; otherwise the magnetic field
would prevent concentration.

Thus I suggest that core formation is essentially the initial stages
of gravitational collapse, at least in Taurus.  Supersonic
flows, however induced, make filaments in this picture.
In the post-shock gas within filaments, gravitational fragmentation and subsequent collapse 
occurs.  This picture predicts that there should be large-scale infall
motions, as seen for example by Tafalla (1998) and Lee, Myers, \&
Tafalla (2001).  It also predicts that
core collapse follows immediately upon core formation (because formation
occurs by gravitational instability); this is also reasonably consistent
with core statistics which suggest that starless cores do not live for
many free-fall times (Jijina et al. 1999).

\section{Core structure}

Most protostellar cores are elongated in
projection.  Myers \etal (1991) argued that typical cores are
more prolate than oblate, which if true would pose significant
problems for simple magnetically-dominated models of core structure,
which most naturally tend to be oblate (flattened along the magnetic
field) than prolate.  Prolate magnetic models have been constructed 
by Curry \& Stahler (2001), but these are quasi-static;
Fiege \& Pudritz (2000) constructed static models, but the plausibility of the
adopted boundary conditions is in doubt.

\begin{figure}[ht]
\plotfiddle{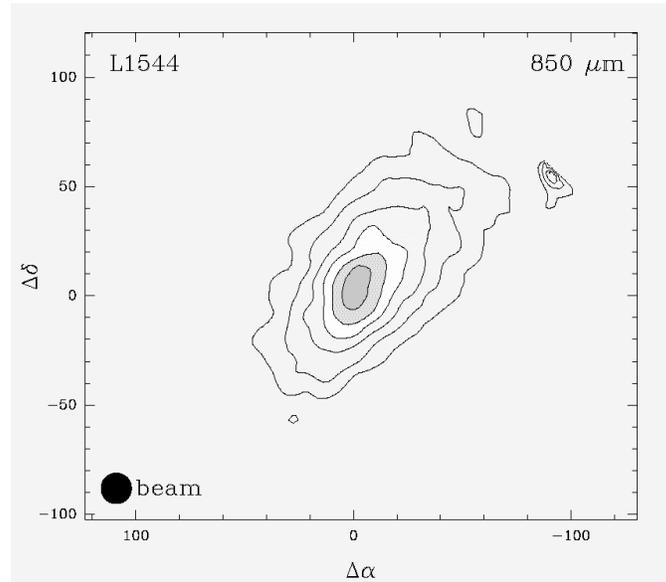}{8.cm}{0}{50}{50}{-150}{-70}
\caption{Submm image of the starless core L1544, 
showing flattened core structure, a flat inner density region,
and an overall asymmetry.  From Shirley \etal 2000).}
\end{figure}

Several independent investigations of core structure have been carried out
which exhibit some tendency to support the notion of prolate (like) cores
(Jones \& Basu 2002; Curry 2002).  However, most studies assume a random distribution
of core inclinations and then make statistical analyses.  It is quite
clear from Figure 3 (right) that the cores in Taurus are NOT randomly
distributed in space; there is a definite overall orientation.
Given the elongation of cores along large-scale filaments,
it seems inescapable that the cores are more prolate-like than oblate. 

In the picture of core formation and evolution advanced in the previous
section, cores are not completely static but are dynamic.  In principle,
this makes it easier to understand prolate and other complex structure;
all forces need not be balanced perfectly, as in true hydrostatic
equilibrium.  Even a relatively ordered, smooth-looking core like L1544
(Figure 4) is not only elongated, but asymmetric along
the major axis; this structure would require some carefully
arranged magnetic field distribution to explain in pure hydrostatic equilibrium.
Such careful arrangements are not necessary in the dynamic model.
Increasingly realistic calculations of dynamic core formation are needed
to test against the observations.  The simulations discussed by
Ballesteros-Paredes et al. (2003) are a start; these calculations ``turn on''
gravity late in the process, which seems unrealistic.  

For purposes of initial analysis, and in the absence of dynamic models, 
static equilibrium models of cores can be useful.  
It has become increasingly popular to compare observations of cores
with so-called Bonnor-Ebert spheres, isothermal, pressure-bounded
models.  The BE sphere models have a significant advantage over the
earlier singular isothermal sphere models; they can be stable, while
the singular sphere is not.  In addition, it has become increasingly
clear that starless cores tend to have flattened density distributions
near their centers (as in L1544, Figure 4) rather than singular distributions
(e.g., Bacmann et al. 2000).  
The BE sphere models can match these properties qualitatively.  In the special
case of B68, a rather spectacular fit can be made to the circularly-averaged
data (Alves et al. 2001).

\begin{figure}[ht]
\plotone{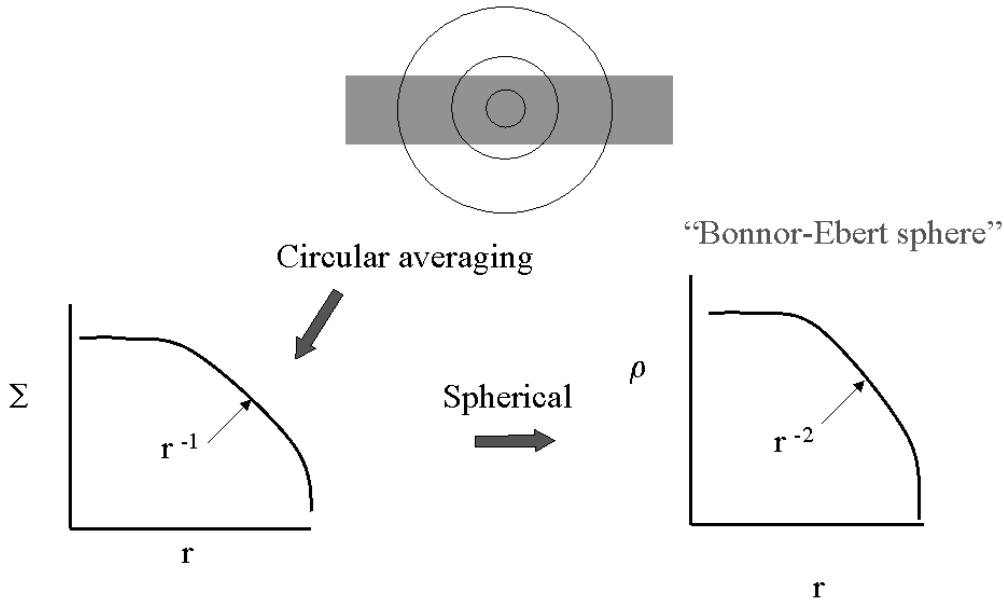}
\caption{Toy model, showing how circular averaging of an intrinsically
elongated surface density distribution, followed by spherical volume
modelling of the density distribution, can yield misleading results
(see text)}
\end{figure}

Caution should be employed that such analyses are not overinterpreted,
especially if used to conclude that cores are generally in hydrostatic
equilibrium, and deriving extremely precise properties.
Averaging can make a very big difference to the interpretation.  For example,
consider the toy core model in Figure 5; it is a uniform rectangle (filament)
projected on the sky.  Now, circularly average; the resulting surface density
has a flat inner region with an $r^{-1}$ falloff, because the mass grows with
$r$ but the surface area increases as $r^2$.  Next, interpret this in terms
of spherical structure; this introduces an extra power of $r$ in the volume
density outside the ``core''.  In this way it is possible to produce something
that looks like a BE sphere, with a flat inner core, and an $r^{-2}$ volume
density dependence outside this core, from a density distribution that is
qualitatively different.

Real cores don't look like my toy core; they are more centrally-condensed.
Nevertheless, it is clear that circular averaging 
smooths over a lot of real structure, and that interpretation in terms
of a spherical structure can be misleading.  I think this is true even in the case
of B68, which is fairly, but not precisely, round; which has significant lumps
in non-smoothed extinction maps; and which has significant, though subsonic,
correlated velocity structure (Lada et al. 2003).  In short, B68 is probably
close to, but not quite in, hydrostatic equilibrium; and B68 is far from typical
of most cores.  We need dynamic models, plus some clever way of applying
these models to non-smoothed data if we are really going to understand cores.

Why is core structure important?  For one thing, the mass infall rate of 
cores with flattened inner structures vary substantially with time,
with higher rates at earlier times (Foster \& Chevalier 1993;
Henriksen, Andr\'e, \& Bontemps 1997; Whitworth \& Ward-Thompson 2001); 
this probably helps to explain
the Class 0 phase of protostellar collapse as a high-infall phase compared
with Class I objects (Andr\'e, Ward-Thompson, \& Barsony 1993).  In contrast,
the singular isothermal sphere models produce a constant infall rate with time.
Perhaps more importantly, cores with flattened inner density distributions
and with strong non-axisymmetric structure are more favorable for the
production of binaries and multiple systems through post-core-collapse 
fragmentation (Bodenheimer et al. 2000, and references therein).

Clearly we are at a very early stage in understanding the consequences
of a more dynamic picture of low-mass core collapse and star formation.
Much more realistic time-dependent numerical simulations are needed to
fully exploit the increased observational capabilities at nearly all wavelengths
to understand star formation and the origin of the stellar initial mass function.

\acknowledgments
I wish to thank Javier Ballesteros and Ted Bergin for their contributions
to the development of these ideas.
This work was supported in part by NASA grant NAG5-9670.


\begin{references}
\reference Alves, J.~F., Lada, C.~J., \& Lada, E.~A.\ 2001, Nature, 409, 159 
\reference Andre, P., Ward-Thompson, D., \& Barsony, M.\ 1993, \apj, 406, 122 
\reference Arons, J.~\& Max, C.~E.\ 1975, \apjl, 196, L77 
\reference Bacmann, A., Andr{\' e}, P., Puget, J.-L., 
Abergel, A., Bontemps, S., \& Ward-Thompson, D.\ 2000, \aap, 361, 555 
\reference Ballesteros-Paredes, J., Hartmann, L., \&
Vazquez-Semadeni, E. 1999, \apj, 527, 285 
\reference Ballesteros-Paredes, J., 
Klessen, R.~S., \& V{\' a}zquez-Semadeni, E.\ 2003, \apj, 592, 188 
\reference Blitz, L.~\& Shu, F.~H.\ 1980, \apj, 238, 148 
\reference{} Bodenheimer, P., Burkert, A., Klein, R.~I., \& Boss, A.~P.\ 2000,
in Protostars and Planets IV, eds. Mannings, V., Boss, A.~P., \& Russell, S.~S.,
University of Arizona Press, Tucson, 675
\reference Bourke, T.L., Myers, P.C., Robinson, G., \& Hyland,
A.R. 2001, ApJ, 554, 916
\reference Crutcher, R.M. 1999, \apj, 520, 706
\reference Curry, C.L. 2002, \apj, in press (astro-ph/0206311)
\reference Curry, C.~L.~\& Stahler, S.~W.\ 2001, \apj, 555, 160
\reference de Bruijne, J.H.J. 1999, \mnras, 310, 585
\reference Elmegreen, B.G.\ 2000, \apj, 530, 277
\reference Fiege, J.~D.~\& Pudritz, R.~E.\ 2000, \apj, 534, 291 
\reference Foster, P.~N.~\& Chevalier, R.~A.\ 1993, \apj, 416, 303 
\reference Franco, J.~\& Cox, D.~P.\ 1986, \pasp, 98, 1076
\reference Henriksen, R., Andre, P., \& Bontemps, S.\ 1997, \aap, 323, 549 
\reference Hartmann, L. 1998, Accretion Processes in Star Formation
(Cambridge University Press), 33
\reference Hartmann, L. 2002, \apj, 578, 914
\reference Hartmann, L., Ballesteros-Paredes, J., \& Bergin, E. 2001,
\apj, 562, 852
\reference Jijina, J., Myers, P.~C., \& Adams, F.~C.\ 1999, \apjs, 125, 161 
\reference Jones, C.~E.~\& Basu, S.\ 2002, \apj, 569, 280 
\reference Klessen, R.S. 2001, \apj, 556, 837
\reference Klessen, R.~S.~\& Burkert, A.\ 2001, \apj, 549, 386
\reference Lada, C.~J., Bergin, E.~A., Alves, 
J.~F., \& Huard, T.~L.\ 2003, \apj, 586, 286 
\reference Larson, R.~B.\ 1981, \mnras, 194, 809 
\reference Larson, R.B. 1985, \mnras, 214, 379
\reference Lee, C.W. \& Myers, P.C. 1999, \apjs, 123, 233
\reference Lee, C.~W., Myers, P.~C., \& Tafalla, M.\ 2001, \apjs, 136, 703
\reference Mac Low, M.-M. 1999, \apj, 524, 169
\reference Mac Low, M.-M., Klessen, R. S., Burkert, A., \& Smith,
M. D. 1998, Phys. Rev. Lett., 80, 275
\reference McKee, C.~F.\ 1989, \apj, 345, 782 
\reference Mestel, L. 1985, in Protostars and Planets II,
eds. D.C. Black \& M.S. Matthews (Tucson: University of Arizona
Press), 320
\reference Mestel, L.~\& Spitzer, L.\ 1956, \mnras, 116, 503 
\reference Miyama, S.M., Narita, S., \& Hayashi, C. 1987a,b, Prog. Theoretical Physics, 78, 1051, 1273
\reference Mizuno, A., Onishi, T., Yonekura, Y., Nagahama, T.,
Ogawa, H., \& Fukui, Y.\ 1995, \apjl, 445, L161
\reference Myers, P.~C., Fuller, G.~A., Goodman, A.~A., \& Benson, P.~J.\ 1991, \apj, 376, 561
\reference Nakano, T.\ 1998, \apj, 494, 587 
\reference Norman, C.~\& Silk, J.\ 1980, \apj, 238, 158 
\reference Padoan, P. \& Nordlund, {\AA}ke 1999, \apj, 526, 279
\reference Passot, T., Vazquez-Semadeni, E., \& Pouquet, A. 1995,
\apj, 455, 536 
\reference Shirley, Y.~L., Evans, N.~J., Rawlings, J.~M.~C., \& Gregersen, E.~M.\ 
2000, \apjs, 131, 249 
\reference Shu, F.H., Adams, F.C., \& Lizano, S. 1987, ARAA, 25, 23
\reference Solomon, P.~M., Sanders, D.~B., \& Scoville, N.~Z.\ 1979, in 
The Large-Scale Characteristics of the Galaxy, 
IAU Symp.~ 84, ed. W.B. Burton (Dordrecht:Reidel), 35 
\reference Stone, J. M., Ostriker, E. C., \& Gammie, C. F. 1998,
ApJ, 508, L99
\reference Tafalla, M., Mardones, D., Myers, P.~C., Caselli, P.,
Bachiller, R., \& Benson, P.~J.\ 1998, \apj, 504, 900
\reference Whitworth, 
A.~P.~\& Ward-Thompson, D.\ 2001, \apj, 547, 317 
\reference Zuckerman, B.~\& Evans, N.~J.\ 1974, \apjl, 192, L149 

\end{references}
\end{document}